\documentclass[%
 reprint,
 amsmath,amssymb,
 aps,
prl,
superscriptaddress
]{revtex4-2}
\usepackage{fancyhdr}
\usepackage{amssymb}
\usepackage{graphicx}
\usepackage{dcolumn}
\usepackage{bm}
\usepackage{color}
\usepackage{array}
\usepackage{etoolbox}

\usepackage{booktabs}
\usepackage{lineno}
\DeclareMathSizes{10}{10}{5.5}{5}

\renewcommand{\footnotetext}

\begin{document}
\title{Integrated quantum communication network and vibration sensing in optical fibers}

\author{Shuaishuai Liu}
\affiliation{State Key Laboratory of Quantum Optics and Quantum Optics Devices, Institute of Opto-Electronics, Shanxi University, Taiyuan 030006, China}
\affiliation{Collaborative Innovation Center of Extreme Optics, Shanxi University, Taiyuan 030006, China}
\author{Yan Tian}
\affiliation{State Key Laboratory of Quantum Optics and Quantum Optics Devices, Institute of Opto-Electronics, Shanxi University, Taiyuan 030006, China}
\affiliation{Collaborative Innovation Center of Extreme Optics, Shanxi University, Taiyuan 030006, China}
\author{Yu Zhang}
\affiliation{State Key Laboratory of Quantum Optics and Quantum Optics Devices, Institute of Opto-Electronics, Shanxi University, Taiyuan 030006, China}
\affiliation{Collaborative Innovation Center of Extreme Optics, Shanxi University, Taiyuan 030006, China}
\author{Zhenguo Lu}
\affiliation{State Key Laboratory of Quantum Optics and Quantum Optics Devices, Institute of Opto-Electronics, Shanxi University, Taiyuan 030006, China}
\affiliation{Collaborative Innovation Center of Extreme Optics, Shanxi University, Taiyuan 030006, China}
\author{Xuyang Wang}
\affiliation{State Key Laboratory of Quantum Optics and Quantum Optics Devices, Institute of Opto-Electronics, Shanxi University, Taiyuan 030006, China}
\affiliation{Collaborative Innovation Center of Extreme Optics, Shanxi University, Taiyuan 030006, China}
\affiliation{Hefei National Laboratory, Hefei 230088, China}
\author{Yongmin Li}
\email{}
\affiliation{State Key Laboratory of Quantum Optics and Quantum Optics Devices, Institute of Opto-Electronics, Shanxi University, Taiyuan 030006, China}
\affiliation{Collaborative Innovation Center of Extreme Optics, Shanxi University, Taiyuan 030006, China}   
\affiliation{Hefei National Laboratory, Hefei 230088, China}

\begin{abstract}
Communication and sensing technology play a significant role in various aspects of modern society. A seamless combination of the communication and the sensing systems is desired and have attracted great interests in recent years. Here, we propose and demonstrate a network architecture that integrating the downstream quantum access network (DQAN) and vibration sensing in optical fibers. By encoding the key information of eight users simultaneously on the sidemode quantum states of a single laser source and successively separating them by a filter network, we achieve a secure and efficient DQAN with an average key rate of $1.88\times {{10}^{4}}$ bits per second over an 80 km single-mode fiber. Meanwhile, the vibration location with spatial resolution of 120 m, 24 m, and 8 m at vibration frequencies of 100 Hz, 1 kHz, and 10 kHz, respectively, is implemented with the existing infrastructure of the DQAN system. Our integrated architecture provides a viable and cost-effective solution for building a secure quantum communication sensor network, and open the way for expanding the functionality of quantum communication networks.	
\end{abstract}

\maketitle
\thanks\email{$^{\ast}$email: yongmin@sxu.edu.cn}

\section{Introduction}
Quantum communication employs quantum states as the carrier of quantum or classical information, plays a significant role in various application scenarios such as quantum key distribution (QKD) \cite{Xu,Advances,Portmann}, quantum teleportation \cite{Pirandola}, quantum digital signatures \cite{Richter}, quantum secret sharing \cite{Richter,Liu}, and quantum e-commerce \cite{Cao}. Although great breakthroughs have been made, traditional QKD is limited to point-to-point communication. Quantum networks can break this limitation and extend the two-user communication to multiple users scenarios. In this direction, simple and cost-effective quantum network architectures as well as the versatile features, are crucial to widespread application of future quantum networks.\\
\indent So far, the demonstrated quantum networks can be classified into fully connected network architectures \cite{Wengerowsky,Joshi} and node-based network architectures \cite{Cao,Richter,Fu-Long,Grasselli,Zhao,Peev,Stucki,Sasaki,Chen-T,Chen,Choi,quantum-access,Huang-D,Park,Brunner,round-trip,Liu,Wang} based on the network topology. The former utilizes untrusted entangled sources and dense wavelength division multiplexing to fully connect all users. Although this architecture is robust and versatile, the user amount $n$ is limited by the number of wavelength channels that is determined by the permutation $A_{n}^{2}$. The latter is subdivided into untrusted relay node network \cite{Fu-Long,Grasselli,Zhao}, trusted relay node network \cite{Peev,Stucki,Sasaki,Chen}, and one-point-to-multipoint centralized network \cite{Cao,Richter,Chen-T,Choi,quantum-access,Huang-D,Park,Brunner,round-trip,Liu,Wang} depending on the type of node. The untrusted relay node network employs remote single-photon interferometry or Bell state measurements to correlate the user information while allowing nodes to be untrusted. It is currently limited to the implementation of network protocols of quantum conference key agreement \cite{Grasselli,Zhao} or quantum secret sharing \cite{Fu-Long}. The trusted relay node network is a fully connected point-to-point QKD link, which can build large-scale network if each node is fully trusted. The one-point-to-multipoint centralized network architecture allows the central node to communicate with multiple nodes to achieve batch sending or receiving of the secret key information. This architecture can share the laser source and the coding system at the transmitter, or the detector and data acquisition system at the receiver. Therefore, it can simplify the configuration and management of the quantum network and reduces the system's construction and maintenance costs. Furthermore, even if a user node fails or disconnects, other nodes can still send or receive information normally.\\
\indent The quantum access networks (QAN) are instrumental to expand the number of users. Compared to downstream QAN (DQAN), upstream QAN (UQAN) usually adopts a time-division multiplexing scheme, which divides the entire communication time into a number of time slots and precisely assigns the corresponding time slots to different users according to predefined rules. The clock synchronization of all users is a prerequisite to ensure smooth system operation, and it is a complex and challenging task. Another scheme is based on active control optical switch, which can route any user to the central node through the optical switch according to the user request. However, as user amount increases, the communication efficiency decreases dramatically.\\
\begin{figure*}
	\centering
	\includegraphics[width=7in]{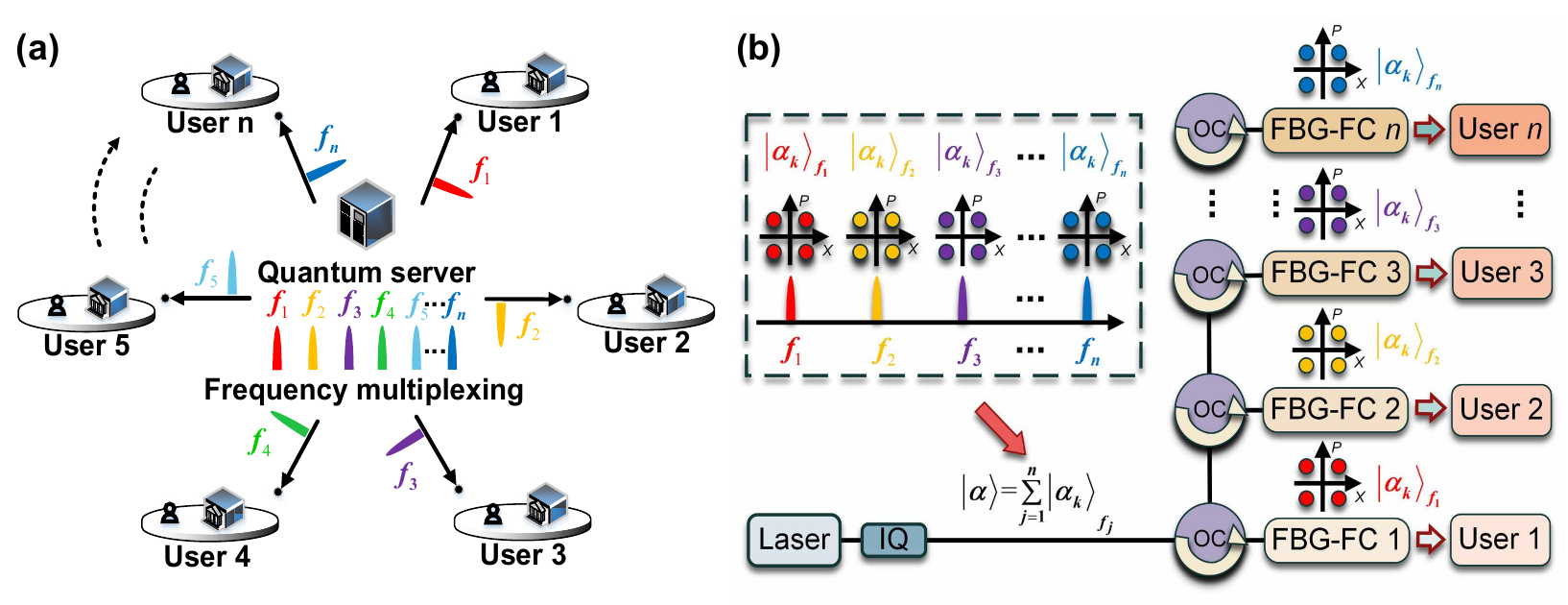}
	\caption{\label{The_downstream_quantum_access_network}\textbf{The schematic diagram of the DQAN.} \textbf{a} The DQAN architecture. The quantum server uses a coding system to prepare multiple sidemode quantum states at sidebands of a single laser source, which are separated one by one using our self-designed filtering network, and sent to the users over the quantum channels. The users measure and extract the key information located at pre-assigned sideband frequencies via the heterodyne detection. \textbf{b} The coding and filtering network. IQ, in-phase and quadrature modulator; OC, Optical circulator; FBG-FC, fiber bragger grating based fiber cavity.}
\end{figure*}
\indent Optical fibers are dominant transmission medium in modern telecommunication networks due to advantages of high bandwidth, low loss, and anti-electromagnetic interference. In addition to transmitting information, significant breakthroughs have been made in forward fiber sensing technology \cite{Marra,Zhan-z,Cantono,Chen-J-P,Ip,Zeng}, which enables monitoring and measurement of seismic activity and tsunamis by analyzing the changes in phase and polarization state of the forward-transmitted light field \cite{Marra,Zhan-z}. Unlike distributed acoustic sensing technology that relies on Rayleigh backscattering in optical fibers \cite{Jousset,He}, the forward fiber sensing technology can achieve much longer sensing distance \cite{Cantono}.\\
\indent Making the quantum communication network systems capable of sensing the ambient environments, namely, building a secure quantum communication sensor network is attractive \cite{Chen-J-P}. This versatile system can expand the functionality of quantum communication networks, and promote the joint development of quantum communication and sensing technology. To this end, it is crucial to introduce the sensing capabilities in the quantum communication network architecture in a low cost and compatible way. More precisely, a seamless combination is desired, that means the sensing system can operate without modifying the architecture of the existing quantum communication network.\\
\indent In this paper, we propose and experimentally demonstrate a secure and efficient DQAN and verify the feasibility of the integration of the DQAN and vibration sensing in optical fibers. By adopting a multi-sideband modulation technique with a single modulator and a filter network consisting of fiber bragger grating based fiber cavities (FBG-FC), the central node simultaneously encodes the key information of 8 users and then separates and sends the key information to each user. The proposed approaches enable the one-point-to-multipoint simultaneous quantum communication without sacrificing the system performance, as well as greatly suppress the cost of quantum network deployment. We analyze the realistic security of the scheme by considering the excess noises from various parts of the system, in particularly, the imperfect modulation and filtering. The spatial resolution of the vibration localization reaches 120 m and 24 m at vibration frequencies of 100 Hz and 1 kHz, respectively. The added value of vibration sensing allows to early detection of critical events, such as earthquakes, tsunamis, landslides, and optical network surveillance that may damage the optical infrastructure, without resorting to dedicated and expensive instruments.
\section{RESULTS}
\subsection{The downstream quantum access network architecture}
For a DQAN, the quantum server should encode multiple key information to enable the expansion of the user amount, without significantly increasing the number of core devices. In addition, the access performance of each user in DQAN wouldn’t sacrifice with the increasing user amount, ensure a robust and efficient quantum networks. \\
\indent  Figure~\ref{The_downstream_quantum_access_network}a shows our DQAN scheme, which is a star network architecture with the center node being the quantum server and the other nodes being the users, it well meets the above requirements. We employ quadrature phase shift keying (QPSK) discrete-modulation continuous variable (CV) QKD protocol. The quantum server uses multi-sideband modulation techniques to encode $n$ key information at different sidemodes of a single frequency laser source with sideband frequency ${{f}_{j}}$, ${j}\in \left\{ 1,2,\ldots ,n \right\}$. The sidemode coherent states prepared can be written as
\begin{equation}\label{1}
\begin{aligned}
	& \left| \alpha  \right\rangle \text{=}\sum\limits_{j=1}^{n}{{{\left| {{\alpha }_{k}} \right\rangle }_{{{f}_{j}}}}},\text{  } \\ 
	& {{\left| {{\alpha }_{k}} \right\rangle }_{{{f}_{j}}}}={{\alpha }_{{{f}_{j}}}}{{e}^{i2\pi k/4}}, k\in \left\{ 0,1,2,3 \right\}. \\ 
\end{aligned}
\end{equation}
Here each sidemode coherent state at ${{f}_{j}}$ is independently and randomly chosen with an identical probability of ${{p}_{k}}=1/4$. The total modulation variance (normalized to the shot noise units (SNU)) of the quantum server is ${{V}_{M}}=\sum\limits_{j=1}^{n}{2\alpha _{{{f}_{j}}}^{2}}$. A single in-phase and quadrature (IQ) modulator is employed in our system to implement the encoding with single sideband modulation. The residual negative first-order sideband can interfere with the measurement of positive first-order sidebands and results in leakage of key information (see Methods).\\
\indent To separate the encoded quantum states and distribute them to the users. One usually employs one-to-N passive beam splitters or active optical switches to split the information into multiple pieces, or different time slots. Due to the added loss or time multiplexing overhead introduced by above two methods, the performance of the system is compromised. To separate the sidemode quantum states without affecting the system performance, a narrow-band filtering system consisting of FBG-FC and optical circulators are designed, as shown in Fig.~\ref{The_downstream_quantum_access_network}b. There are two criteria for the design of the fiber cavity. Firstly, the free spectral region (FSR) ${{v}_{FSR}}=c/2dl$ should be larger than the frequency range of entire sidemodes, where $c$ is the speed of light in a vacuum, $d$ is the refractive index of the fiber core, and $l$ is the cavity length.  Secondly, the linewidth of the cavity $\Delta v={{v}_{FSR}}/F$ ($F$ is the fineness of the cavity) should be wider than the bandwidth of the each sidemode (determined by the system baud rate and modulation format) and narrower than the frequency spacing of the adjacent sidemodes. Above criteria ensure that each sidemode state can be faithfully separated with the least crosstalk to other sidemodes. The imperfect filtering may introduce excess noises and deteriorate the performance of the system (see Supplementary Note 3 for details). \\
\indent The separated sidemode states are sent to each user over an unsecured quantum channel. After measuring the received signals using heterodyne detection, the frequency and phase of each sidemode signal are recovered with the aid of the accompanying pilot tone. To extract the secret key from the raw data, the quantum server implements the standard secret key rate estimation procedure of the QPSK CV-QKD, reverse data reconciliation, and private amplification.
\subsection{\label{sec:level6}Secret key rate}
The asymptotic secret key rate of each user in DQAN is given by \cite{Ghorai-S,Lin,Lin-J}
\begin{equation}\label{2}
	{{K}^{\infty }}=R\left( \underset{{{\rho }_{AB}}\in \mathcal{S}}{\mathop{\min }}\,D\left[ \mathcal{G}\left( {{\rho }_{AB}} \right)\|\mathcal{Z}\left[ \mathcal{G}\left( {{\rho }_{AB}} \right) \right] \right] \right)-{{p}_{\text{pass}}}{{\delta }_{\text{EC}}},
\end{equation}
where $R$ is the repetition rate the system. $D\left( \rho \left\| \sigma  \right. \right)=\text{Tr}\left( \rho \text{lo}{{\text{g}}_{2}}\rho  \right)-\text{Tr}\left( \rho \text{lo}{{\text{g}}_{2}}\sigma  \right)$, denotes the quantum relative entropy, $\mathcal{G}$ denotes a fully positive and trace non-increasing map for classical post-processing processes, ${{\rho }_{AB}}$ denotes the joint state of the quantum server and the user after transmission through the quantum channel. $\mathcal{Z}$ describes a pinching quantum channel for obtaining key mapping results, ${{p}_{\text{pass}}}$ denotes the sifting probability that the data from a given round will be used to generate the key. ${{\delta }_{\text{EC}}}$ denotes the leakage information of each signal pulse during error correction. \\
\indent During the actual implementation of the DQAN, the imperfections of the devices can lead to potential security loopholes, which can be exploited by Eve \cite{Qin-H,Li-C}. In our scheme, the IQ modulator works in a weak modulation regime and only the positive first-order modulation sideband is employed. However, the residual mirror sideband due to finite sideband suppression ratio may leak some of the key information. Furthermore, the finite isolation degree of the filtering network that separates the different sidemodes can mix the signal fields of other users into the current user and cause information leakage. \\
\indent To close above security loopholes, we establish the security analysis model of the DQAN under the imperfect modulation and filtering attacks (See “Methods” for the detailed analysis). Based on the results of the security analysis, one of the constraints in minimizing the quantum relative entropy using convex optimization should be modified to
\begin{equation}\label{3}
	\begin{aligned}
	& \text{T}{{\text{r}}_{B}}\left[ {{\rho }_{AB}} \right]=\sum\limits_{l,h=0}^{3}{\sqrt{{{p}_{l}}{{p}_{h}}}}\left\langle  {{{{\alpha }'}}_{h}} \right|\left. {{{{\alpha }'}}_{l}} \right\rangle \left| l \right\rangle {{\left\langle  h \right|}_{A}},\\
	& {{{\alpha }'}_{{f}_{j}}}=g{{\alpha }_{{f}_{j}}},\\
	\end{aligned}
\end{equation}
where $g=1/({{g}_{1}}\sqrt{\text{1}-{{S}_{1}}-{{S}_{2}}})$ is the correction factor due to the imperfect modulation and filtering effect (see Supplementary Note 1 for the detailed analysis).
\begin{figure*}[!htbp]
	\centering
	\includegraphics[width=7in]{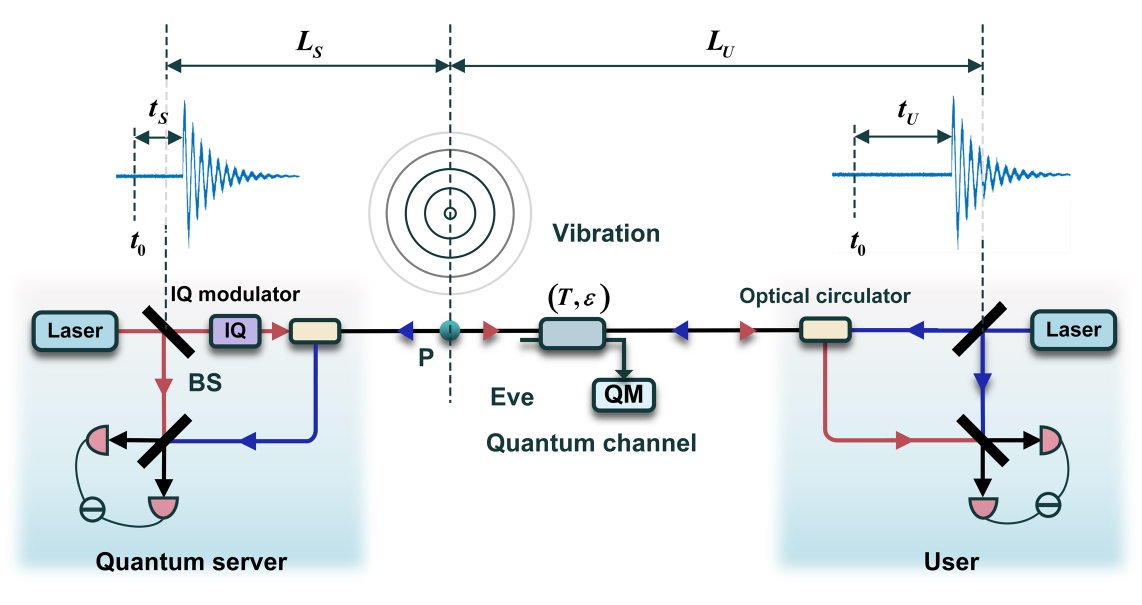}
	\caption{\label{Fiber_vibration_sensing}\textbf{The schematic diagram of the optical fiber vibration sensing using the existing DQAN devices.} The quantum server sends the quantum signals and pilot tones to the user. The user injects a weak probe beam back into the fiber through an optical circulator. Then, they measure and analyze the phase variations of the forward pilot tone and backward probe beam to achieve fiber vibrational localization through heterodyne detection and digital signal processing.}
\end{figure*}
\subsection{Optical fiber vibration sensing}
Mechanical vibrations acting on the optical fiber can cause changes in the strain and the refractive index of the fiber core, which further induce phase variations of the transmitting light. In our DQAN, the users employ the pilot tones to estimate the frequency offset and phase drift of the transmitting signal light from the quantum server. In this case the acquired phase variation information can be exactly applied to monitor the vibration acting on the optical fibers in real time.\\
\indent To achieve vibration localization, two-way forward transmission is required (Fig.~\ref{Fiber_vibration_sensing}). To this end, the users split a small portion of the laser and inject it into the quantum channel via a fiber circulator. The quantum server measure the transmitting light from the users using heterodyne detection and estimate the phase variation information. The measured phase signals of the users and quantum server can be given by \cite{Ip}
\begin{equation}\label{4}
\begin{aligned}
	& \Delta {{\phi }_{S}}(t)=2\pi \left( {{f}_{U}}-{{f}_{S}} \right)t+\left[ \phi _{U}^{laser}\left( t-\tau  \right)-\phi _{S}^{laser}(t) \right] \\ 
	& \qquad \qquad \; +\phi _{vib}\left( t-\tau_{S}  \right)+\phi _{S}^{sys}(t), \\ 
\end{aligned}
\end{equation}
\begin{equation}\label{5}
	\begin{aligned}
		& \Delta {{\phi }_{U}}(t)=2\pi \left( {{f}_{S}}-{{f}_{U}} \right)t+\left[ \phi _{S}^{laser}\left( t-\tau  \right)-\phi _{U}^{laser}(t) \right] \\ 
		& \qquad \qquad \; +\phi _{vib}\left( t-\tau_{U}  \right)+\phi _{U}^{sys}(t), \\ 
	\end{aligned}
\end{equation}
where ${{f}_{S}}$, ${{f}_{U}}$ and $\phi _{S}^{laser}$, $\phi _{U}^{laser}$ denote the laser frequency and phase of the quantum server and the users respectively. $\phi _{vib}$ is the phase fluctuations arising from the fiber link. $\tau_{S} ={L_{S}}/{\left( {c}/{d}\right)}$, $\tau_{U} ={L_{U}}/{\left( {c}/{d}\right)}$, and $\tau ={L}/{\left( {c}/{d}\right)}$, where $L=L_{S}+L_{U}$, $L_{S}$ and $L_{U}$ are the distances from the vibration point to the quantum server and the user, respectively. $\phi _{S}^{sys}$ and $\phi _{U}^{sys}$ are the equivalent (broadband) phase noise introduced by the receiver and the fiber link. \\
\indent To locate the vibration source at certain frequency band, the first frequency offset terms in Eqs. (\ref{4}) and (\ref{5}) are compensated by utilizing frequency estimation. Next, digital band-pass filters are employed to suppress the laser phase and system noises and improve the signal to noise ratio. Then the delay difference $\Delta t\text{=}{{t}_{S}}-{{t}_{U}}$ is tracked by correlating the filtered phase signals $\Delta {{\phi }_{S,F}}(\text{t})$ and $\Delta {{\phi }_{U,F}}(\text{t})$ and finding the peak value of the correlation coefficients, here ${{t}_{S}}$ and ${{t}_{U}}$ are the propagation delays from the vibration source to the quantum server and the user, respectively. At this stage, the vibration position is given by
\begin{equation}\label{6}
	{{L}_{S}}=\frac{1}{2}\left( L-\frac{c\Delta t}{d} \right).
\end{equation}
\indent In above, we have shown that our scheme can support both the DQAN and the vibrational sensing without modifying the existing hardware architecture and only an additional heterodyne detector is required.
\begin{figure*}[!t]
	\centering
	\includegraphics[width=7.2in]{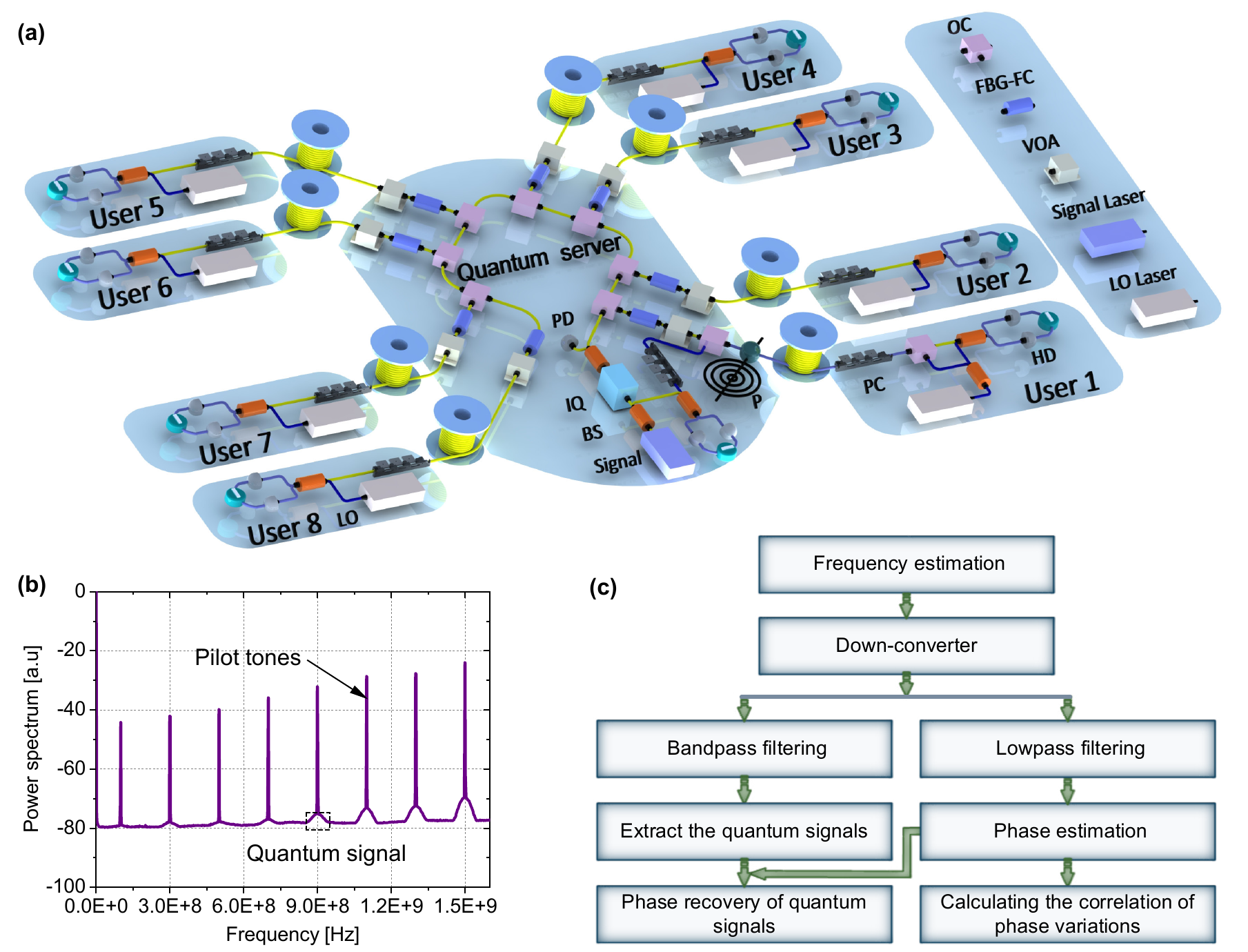}
	\caption{\label{Experimental_setup_and_principles}\textbf{ Experimental diagram of the integrated DQAN and fiber vibration sensing.} \textbf{a} Experimental setup. BS, beam splitter; IQ, in-phase and quadrature modulator; PD, photoelectric detector; P, vibration position; PC, polarization controller; HD, heterodyne detector; OC, Optical circulator; FBG-FC, fiber bragger grating based fiber cavity; VOA, variable optical attenuator. \textbf{b} Power spectrum of eight quantum signals and pilot tones. The frequency interval between adjacent quantum signals is 200 MHz. The increasing intensity of the quantum signals and pilot tones with the frequency is due to the compensation of filtering loss. \textbf{c} Digital signal procedures to extract the quantum signals and vibration location.}
\end{figure*}
\subsection{\label{sec:level3} Experimental implementation}
The experimental setup of the integrated DQAN and fiber vibration sensing is shown in Fig.~\ref{Experimental_setup_and_principles}a. The DQAN consists of a quantum server and eight users connected via standard single-mode fiber spools. The quantum server performs multi-sideband modulation on a 100 Hz linewidth continuous-wave single-frequency laser (NKT-X15) by an IQ modulator (ixBlue). The baud rate of the quantum symbols at each sideband is 50 MHz and the center frequency of the quantum signals is $F\text{=}A\text{+}j\Delta {{\Omega }_{s}}$, $j\in \left( 0,1,\ldots ,7 \right)$, where $A=100$ MHz and $\Delta {{\Omega }_{s}}=200$ MHz. The pilot tones that providing frequency and phase reference for the quantum signals are frequency multiplexed at the center frequency of the quantum signals. An arbitrary waveform generator (AWG) with a 10 Gsamples/s sampling rate converts the generated digital signals into analogue signals to drive the IQ modulator. The multi-sideband quantum signals are separated through the self-designed filtering network and then sent to the users through the quantum channels. \\
\begin{table*}[!htbp]
	\caption{\label{table_1}\textbf{The parameters of the experimental system.} $F$, center frequency of the quantum signals; $V_\mathrm{A}$, modulation variance of the quantum signals; $v_\mathrm{el}$, electronic noise of the heterodyne detection; $\eta$, detection efficiency; $\beta$, reconciliation efficiency; $\varepsilon$, excess noise; $T$, transmittance of the quantum channel; $K$, secret key rate. }
	\resizebox{\textwidth}{20mm}{
		\begin{tabular}{ccccccccccc}
			\hline\hline
			Users&$F$ (MHz)&$V_\mathrm{A}$ (SNU)&${{v}_\mathrm{el}}$ (SNU) &$\eta$ (\%)&$\beta$ (\%)&$\varepsilon$ (SNU)&$T$ & $K$ (bits per second) \\ \hline
			User 1&100&1.16&0.15&45&95&0.022& 0.024 &$1.59\times {{10}^{4}}$\\
			User 2&300&1.18&0.17&53&95&0.020 & 0.025 &$2.28\times {{10}^{4}}$ \\
			User 3&500&1.17&0.21&50&95&0.021& 0.025&$2.03\times {{10}^{4}}$\\
			User 4&700&1.17&0.29&52&95&0.020& 0.026&$2.17\times {{10}^{4}}$\\
			User 5&900&1.17&0.19&51&95&0.022 & 0.025 &$1.76\times {{10}^{4}}$ \\
			User 6&1100&1.21&0.17&54&95&0.021& 0.024 &$1.93\times {{10}^{4}}$\\
			User 7&1300&1.14&0.16&51&95&0.024& 0.025 &$1.55\times {{10}^{4}}$\\
			User 8&1500&1.17&0.16&54&95&0.023& 0.024&$1.74\times {{10}^{4}}$ \\  \hline
	\end{tabular}}
\end{table*}
\begin{figure*}[!htbp]
	\centering
	\includegraphics[width=5.6in]{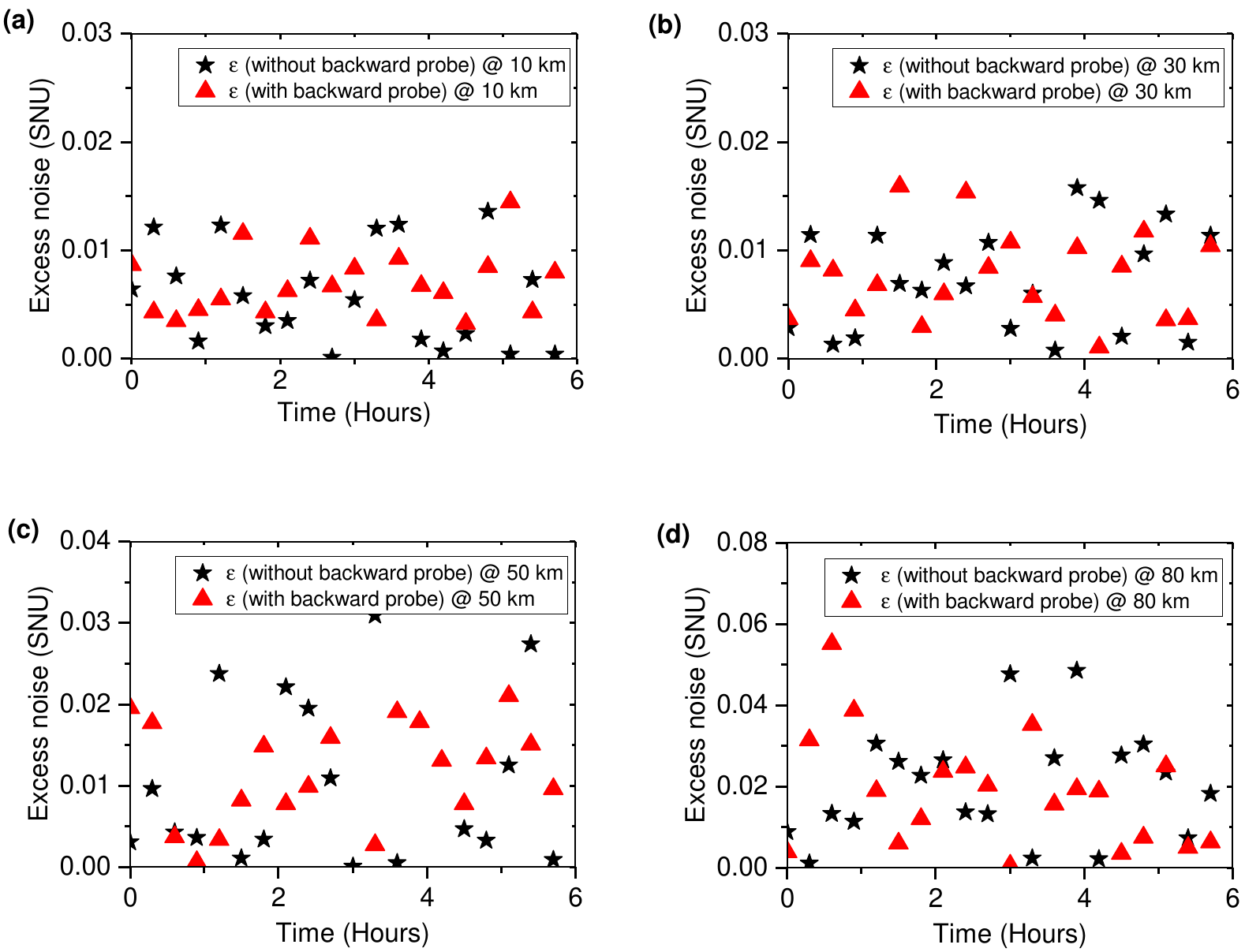}
	\caption{\label{The_excess_noise} \textbf{The excess noise of User 1 with and without the backward probe beam.}\textbf{ a, b, c, d} The excess noise at 10, 30, 50, 80 km of single-mode fibers. The black pentagram denotes the excess noise without the backward probe beam. The red triangle denotes the excess noise with the backward probe beam.}
\end{figure*}
\indent Figure~\ref{Experimental_setup_and_principles}b depicts the measured power spectrum of the quantum signals and pilot tones. The resonance peak of the FBG-FC is precisely tuned to the center frequency of each sidemode using a piezoelectric transducer (PZT). The temperature of the FBG-FC is stabilized at room temperature with an accuracy less than 0.01°C to prevent temperature drift, which would cause the frequency shift of the resonance peak of the FBG-FC and induce the intensity fluctuations of the quantum signals.\\
\indent To investigate the fiber vibration sensing capability of our setup, we suspend 1.5 m of the single mode fiber in quantum channel over two fixed iron rods and simulate the vibration of outdoor suspension fiber cables by manually tapping on the rods. For the high frequency vibration simulation above 1 kHz, we install program-controlled PZT in the quantum channel (Fig.~\ref{Experimental_setup_and_principles}a), which generates sinusoidal modulation at 1 kHz and 10 kHz. \\
\indent After receiving the quantum signals, the users recover the polarization states of the signal fields with a manual fiber polarization controller and perform the heterodyne detection with the local local oscillator (LLO) \cite{Qi,Adnan}. The LLO is generated by an independent laser with frequency difference of 50 MHz to the quantum server's laser. A small portion of the laser with power of $-43$ dBm at user 1 is inversely injected into the quantum channel via an optical circulator to provide the probe beam for the fiber vibration sensing. The output signals from the heterodyne detectors at the quantum server and users were synchronously acquired using an oscilloscope with a sampling rate of 6.25 Gsamples/s.\\
\indent The frequency and phase of the pilot tones, and the quantum signals are extracted through a digital signal processing (DSP) module as shown in Fig.~\ref{Experimental_setup_and_principles}c. By using the retrieved relative frequency of the pilot tone and the probe beam, the quantum signals, the pilot tones, and the probe beams are down-converted to the baseband. A digital band-pass filter is employed to extract the quantum signals by filtering out the pilot tones and high frequency noises. The low-pass filtered quadrature signals of the pilot tone and the probe beam are utilized for phase estimation to recover the quantum signals and sense the vibration, respectively. For fiber vibration sensing, the quantum server and the users correlate both the retrieved phase variations of the forward pilot tones and backward probe beams.
\begin{figure}[!htbp]
	\centering
	\includegraphics[width=3.4in]{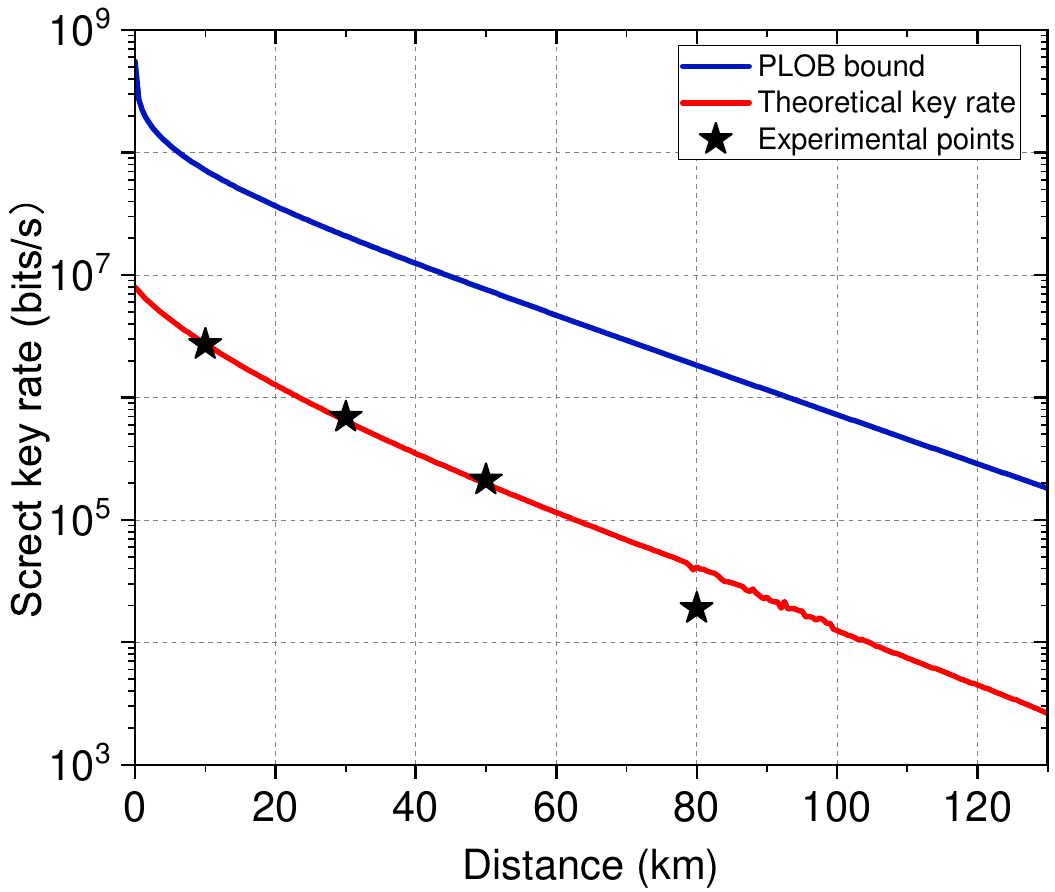}
	\caption{\label{Secret_key_rates} \textbf{Secret key rates of DQAN.}The blue solid line is the PLOB bounds. The red solid line is the simulation key rate calculated by using the experimental parameters. The black pentagrams are the experimental secret key rate (average value for the eight users) over 10, 30, 50, and 80 km single mode fibers, respectively.}
\end{figure}
\begin{figure*}[!htbp]
	\centering
	\includegraphics[width=7in]{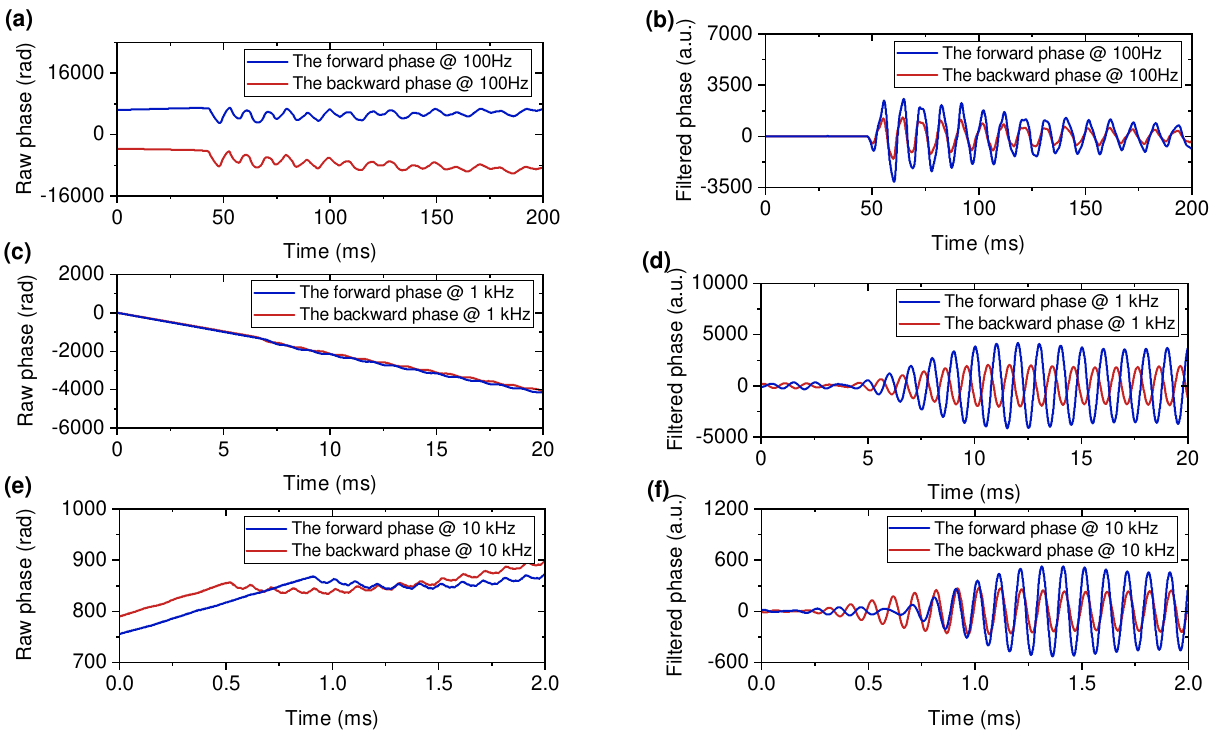}
	\caption{\label{The_phase_perturbations} \textbf{The recovered phases at different vibration frequencies.} The blue and red lines denote the raw phase variations of the forward pilot tone and backward probe beam at vibration frequencies of 100 Hz (\textbf{a}), 1 kHz (\textbf{c}), and 10 kHz (\textbf{e}), respectively. \textbf{b d f} depict the filtered phase variations after digital bandpass filters.}
\end{figure*}
\subsection{\label{sec:leve26}Experimental results}
Table~\ref{table_1} shows the experimental parameters of our integrated DQAN and vibration sensing at 80 km standard single-mode fiber, including the modulation variance, excess noises, channel transmittance, detection efficiency, and secret key rate, etc. The estimation of the secret key rate depends on the first and second moments of the Bob’s quadratures and Alice’s modulation variance. Here the excess noise is estimated by considering the realistic physical channels and used only to indicate the excess noises caused the system itself, which are attributed to the Eve’s attack and critical to the secret key rate (see Supplementary Note 1 for the details). \\
\indent The excess noise in our system mainly consists of the following components
\begin{equation}\label{7}
	\varepsilon \text{=}{{\varepsilon }_{CV-QKD}}+{{\varepsilon }_{\text{SASRS}}}\text{+}{{\varepsilon }_{Freq}}\text{+}{{\varepsilon }_{Filt}},
\end{equation}
 where ${{\varepsilon }_{CV-QKD}}$ is the excess noise in a standard CV-QKD system \cite{Laudenbach}, ${{\varepsilon }_{\text{SASRS}}}$ is the Raman scattering noise due to the backward probe beam, ${{\varepsilon }_{Freq}}$ is the frequency crosstalk noise, and ${{\varepsilon }_{Filt}}$ is the filtering crosstalk noise. Notice that the Rayleigh scattering of the backward probe beam in fiber lies at zero-frequency band that differs from the frequency range of the quantum signals and is eliminated during the demodulation of the quantum signals. Here, we focus on the Raman scattering noise, frequency crosstalk noise, and filtering crosstalk noise (see Supplementary Note 3 for more details). The power of the backward probe beam is around $-43$ dBm in our experiment, which induce a Raman scattering noise level of $4.69\times {{10}^{-6}}$ SNU. To verify this prediction, we measured the excess noises of user 1 with and without the backward probe beam in 10 km, 30 km, 50 km, and 80 km single mode fibers, respectively (Fig.~\ref{The_excess_noise}). The average values of the excess noises with (without) backward probe beam are almost the same: 0.0058 (0.0069), 0.0073 (0.0075), 0.013 (0.013), and 0.020 (0.019). The results indicate that the existence of the backward probe beam has negligible effect on the DQAN system and agrees with the theoretical predictions.\\
 \indent Figure ~\ref{Secret_key_rates} shows the secret key rates of the DQAN system. The black pentagrams denotes the key rate versus the length of the single mode fibers. The red and blue lines denote the theoretical key rate using the experimental parameters and the PLOB bound \cite{Pirandola-S}, respectively. The experimental secret key rates are $2.71\times {{10}^{6}}$ bps (10 km), $6.92\times {{10}^{5}}$ bps (30 km), $2.12\times {{10}^{5}}$ bps (50 km), and $1.88\times {{10}^{4}}$ bps (80 km), respectively.\\
 \indent Figure~\ref{The_phase_perturbations} shows the recovered phases of the forward pilot tone and backward probe beam at different vibration frequencies. To achieve high resolution positioning, it is critical to select a digital bandpass filter with appropriate type, bandwidth and order, to improve the signal to noise ratio of the raw phases. For the vibration frequency at 100 Hz, we use a high-precision infinite impulse response (IIR) Butterworth filter. In contrast, two cascaded finite impulse response (FIR) windows filters are employed for the filtering of the raw phases at 1 kHz and 10 kHz (see Supplementary Note 2 for details). The relative delay of the vibration signals measured by the quantum server and the users is obtained by calculating the correlation between the two extracted phase variations, and the vibration position is successively determined by Eq. (\ref{6}). The spatial resolution of the vibration location is 120 m, 24 m, and 8 m at vibration frequencies of 100 Hz, 1 kHz, and 10 kHz, respectively. 
\subsection{Discussion}
In our proof-of-principle experiment, eight users share the same local laser to provide their LO. So that each user require a high bandwidth heterodyne detector that can cover the frequency range of the respective sidemode and the sampling rate of the analog to digital convertor (ADC) increases with the frequency of the sidemodes. In real-world network applications, where each user has its own local laser and its frequency can be tuned to close to that of the received sidemodes. In this case, the quantum signals are shifted down to low frequency range, so that low bandwidth heterodyne detectors and low ADC sampling rate are sufficient to acquire the quantum signals. This feature significantly simplifies the measurement apparatus and reduce the cost of the DQAN. Moreover, our scheme show high potential to be extended to more users if a higher bandwidth modulator are employed, where very high bandwidths over 100 GHz have been demonstrated in a thin-film lithium niobate platform \cite{Wang-C}.\\
\indent  In our current experiment, a single-point vibration detection and location is demonstrated for one user. It is possible to achieve distributed sensing of multi-point vibration by using advanced data processing methods such as Kalman filtering, and machine learning. On the other hand, single user vibration sensing can pinpoint the vibration occurring exactly on or very close to the fiber. If multiple users collaborate, the vibration signals far from the fiber can be located. For instance, the earthquake epicenter can be located by using two bidirectional fiber links \cite{Marra}. More precisely, the coordinates of the epicenter is determined from the location point of first arriving of the seismic wave along the fiber and simple geometry. Furthermore, a remote timing synchrony between the forward and backward receivers is essential in practical applications, which can be achieved by locking to GPS.\\
\indent In conclusion, we have proposed and demonstrated a scheme that integrating DQANs and fiber vibration sensing. The scheme requires only one transmitter to encode the key information and can achieve simultaneous communication between the quantum server and the multiple users. By using the existing infrastructure of the DQAN system and introducing only one heterodyne detector at the quantum server, the function of fiber vibration detection and localization can be achieved without adverse effects on the DQAN. We also analyzed the information leakage by imperfect modulation and quantum signal filtering and proposed the corresponding countermeasures. Our results provide a new network architecture for future quantum network deployments as well as opening a new door for multi-functional quantum networks.
\section{Methods}
The imperfections in realistic physical devices of the integrated DQAN and vibration sensing system may suffer from Eve's attack. In this part, we analyze the practical security of our system under the imperfect modulation and filtering and propose the corresponding countermeasures.
\subsection{The negative sideband}
For ideal carrier suppression single sideband modulation (CS-SSB) modulation, the negative sideband is absent. In practice, the negative sidebands cannot be completely suppressed due to the imprecise locking of the bias points, the modulation imbalance of the IQ modulator, and so on. Eve can extract the key information by acquiring the negative sidebands and raise security threats.\\
\indent In general, the modulation depths of the IQ modulator are $\mu _{j}^{I}\text{=}{{\mu }_{I}}=\mu +\sigma$ and $\mu _{j}^{Q}\text{=}{{\mu }_{Q}}=\mu -\sigma$, respectively, where $\sigma$ is a small value that characterize the modulation imbalance of the I and Q signals. By inserting the expressions of modulation depths into the output field of the IQ modulator, we have
\begin{widetext}
	\begin{equation}\label{8}
	{{E}_{out}}\left( t \right)={{E}_{in}}\left( t \right)\left[ \sin \left( \sum\limits_{j=1}^{n}{\mu _{j}^{I}\cos \left( {{\omega }_{j}}t \right)} \right)+\sin \left( \sum\limits_{j=1}^{n}{\mu _{j}^{Q}\sin \left( {{\omega }_{j}}t \right)} \right){{e}^{i\pi \frac{{{V}_{Pb}}}{2{{V}_{{\pi }/{2}\;}}}}} \right],
	\end{equation}
\end{widetext}
where ${{E}_{in}}\left( t \right)\text{=}E_{0}{{\text{e}}^{i{{\omega }_{s}}t}}$ is the input optical field into the IQ modulator. ${{V}_{Pb}}$ is the bias voltage between the I and Q path. After some algebras \cite{Jain-N,Wang-H}, Eq. (\ref{8}) can be rewritten as
\begin{widetext}
	\begin{equation}\label{9}
		\begin{aligned}
&{{E}_{out}}\left( t \right)\approx {{E}_{in}}\left( t \right)\sum\limits_{m=1}^{\infty }{\left( J_{0}^{j-1}({{\mu }_{I}}){{J}_{2m-1}}({{\mu }_{I}}){{(-1)}^{m+1}}\sum\limits_{j=1}^{n}{\text{cos}\left[ (2m-1){{\omega }_{j}}t \right]}-iJ_{0}^{j-1}({{\mu }_{Q}}){{J}_{2m-1}}({{\mu }_{Q}})\sum\limits_{j=1}^{n}{\sin \left[ (2m-1){{\omega }_{j}}t \right]} \right)}\\
&\qquad \quad \; \text{=}E_{0}\sum\limits_{j=1}^{n}{\left[ \frac{\mu }{2}{{\text{e}}^{i\left( {{\omega }_{s}}+{{\omega }_{j}} \right)t}}\text{+}\frac{\sigma }{2}{{\text{e}}^{i\left( {{\omega }_{s}}-{{\omega }_{j}} \right)t}} \right]},\\
		\end{aligned}
	\end{equation}
\end{widetext}
where ${{J}_{1}}({{\mu }_{I}})\approx E_{0} (\mu +\sigma )/{2}$ and ${{J}_{1}}({{\mu }_{Q}})\approx E_{0}(\mu -\sigma )/{2}$.\\
\indent Equation (\ref{9}) indicates that the IQ modulator produces negative sideband terms. In the experiment, we suppressed the negative sidebands by finely adjusting the phase and amplitude of the IQ modulation signals, and a suppression ratio larger than 35 dB was obtained. Considering that the negative first-order sidebands $E_{0} (\sigma /2){{\text{e}}^{i\left( {{\omega }_{\text{s}}}-{{\omega }_{\text{j}}} \right)t}}$ are the dominant residual terms under the weak modulation condition, the ratio of the positive first-order sideband to the total sidebands is approximated as 
\begin{equation}\label{10}
	{{g}_{1}}\approx \sqrt{\frac{\mu }{\mu +\sigma }}\text{=}\sqrt{\frac{1}{1.000316}}\approx 0.9998.
\end{equation}
\begin{figure*}[!t]
	\centering
	\includegraphics[width=7.2in]{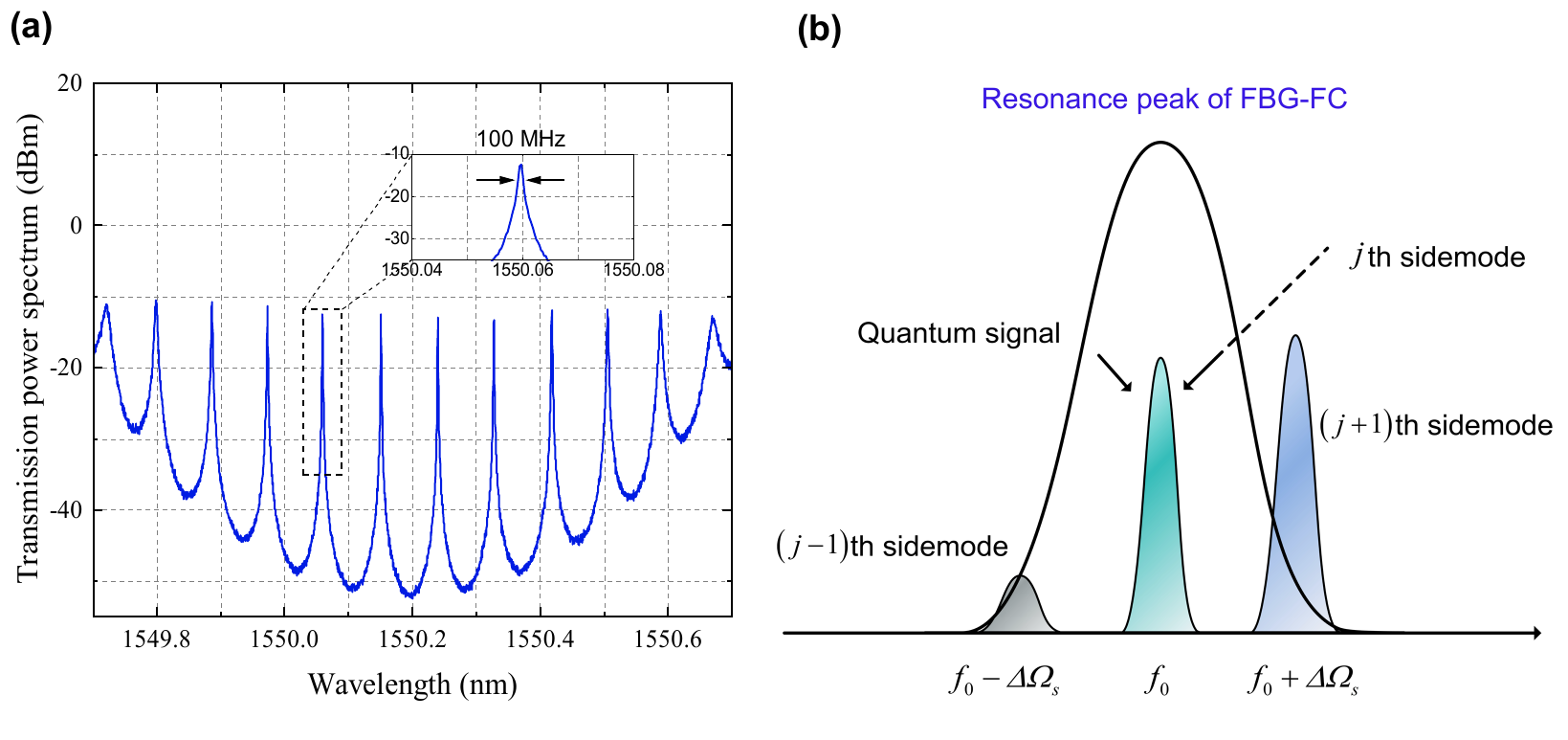}
	\caption{\label{FBG_resonant_peak} \textbf{The schematic diagram of imperfect filtering.} \textbf{a} The transmission power spectrum of FBG-FC. The resonance peak in the dashed box is employed to separate the different sidemodes. Three factors are considered to select the resonance peak: the wavelength of the laser, the extinction ratio of the transmitted peak, and the direction of the resonance wavelength movement when stretching the FBG-FC by PZT. The linewidth of the FBG-FC (Full width at half maximum, FWHM) is around 100 MHz and the frequency interval between the adjacent users is 200 MHz. The finite extinction ratio of the FBG-FC may result in the leakage of the adjacent users' key information. \textbf{b} The leakage due to imperfect filtering. The $(j-1)$th sidemode is the residual sidemode after the sidemode extraction by the $(j-1)$th user.}
\end{figure*}
\subsection{\label{sec:leve21} Imperfect filtering}
Figure~\ref{FBG_resonant_peak}a shows the measured transmission spectrum of the FBG-FC by an optical spectrum analyzer. We select the resonance peak in the black dashed box to separate the different sidemodes. This is determined by the wavelength of the laser, the extinction ratio of the resonance peak, and the direction of the resonance wavelength movement when stretching the FBG-FC by PZT. For a realistic filter with finite extinction ratio, the filter network is unable to fully reject the quantum signals of the other users when selecting the quantum signal of the target user (Fig.~\ref{FBG_resonant_peak}b). The leaked information may open up security loopholes, for example, Eve can cooperate with the dishonest users to perform an attack.\\
\indent The normalized transmission spectrum of the FBG-FC can be expressed as a Lorentzian function
\begin{equation}\label{11}
	T\left( f \right)=\frac{1}{\pi }\frac{\frac{1}{2}\Delta v }{{{\left( f-{{f}_{0}} \right)}^{2}}+{{\left( \frac{1}{2}\Delta v  \right)}^{2}}},
\end{equation}
where ${{f}_{0}}$ is the center frequency of the resonance peak, $\Delta v $ is the FWHM linewidth. In our experiment, the linewidth $\Delta v $ of the FBG-FC is about 100 MHz. We consider only the leaked information from the adjacent users (the leaked information from the non-adjacent users is negligible). From Eq. (\ref{11}), the normalized intensity of the $j$th sidemodes to the $(j-1)$th and $(j+1)$th sidemodes are given by
\begin{equation}\label{12}
	\begin{aligned}
	&{{S}_{j-1}}\text{=}{{R}_{j}}\frac{\int_{{{f}_{0}}-\Delta {{\Omega }_{s}}-\Delta {{f}_{s}}/2}^{{{f}_{0}}-\Delta {{\Omega }_{s}}+\Delta {{f}_{s}}/2}{T\left( f \right)df}}{\int_{{{f}_{0}}-\Delta {{f}_{s}}/2}^{{{f}_{0}}+\Delta {{f}_{s}}/2}{T\left( f \right)df}}=0.0180,\\
	& {{S}_{j+1}}\text{=}\frac{\int_{{{f}_{0}}+\Delta {{\Omega }_{s}}-\Delta {{f}_{s}}/2}^{{{f}_{0}}+\Delta {{\Omega }_{s}}+\Delta {{f}_{s}}/2}{T\left( f \right)df}}{\int_{{{f}_{0}}-\Delta {{f}_{s}}/2}^{{{f}_{0}}+\Delta {{\text{f}}_{s}}/2}{T\left( f \right)df}}=0.0792,\\
	\end{aligned}
\end{equation}
where ${{R}_{j}}=0.227$ denotes the residual reflectivity of the FBG-FC for the $j$th sidemode.
\subsection{\label{sec:leve27}The countermeasures against the negative sideband and nonideal filtering attacks}
In above, we show that the non-ideal CS-SSB modulation can result in the emergence of the negative sidebands. On the other hand, the imperfect filtering can cause the leakage of the sidemodes to the adjacent user channels. To close the security loopholes arising from these two aspects, we can correct the amplitude of prepared coherent states at Alice’s site and rewrite Eq. (\ref{1}) as, 
\begin{equation}\label{13}
	\begin{aligned}
&\left| {{\alpha }'} \right\rangle \text{=}\sum\limits_{j=1}^{n}{{{\left| {{\alpha }_{k}}^{\prime } \right\rangle }_{{{f}_{j}}}}},\text{  }{{\left| {{\alpha }_{k}}^{\prime } \right\rangle }_{{{f}_{j}}}}={{{\alpha }'}_{{{f}_{j}}}}{{e}^{i2\pi k/4}}\text{,   }\\
&{{{\alpha }'}_{{{f}_{j}}}}={{\alpha }_{{{f}_{j}}}}/({{g}_{1}}\sqrt{1-{{S}_{j-1}}-{{S}_{j+1}}}), k\in \left\{ 0,1,2,3 \right\},\\
	\end{aligned}
\end{equation}
where ${{\alpha }_{{{f}_{j}}}}$ is the measured complex amplitude of the coherent states prepared for the $j$th user, and ${{{\alpha }'}_{{{f}_{j}}}}$is the corrected value. The correction here is equivalent to virtually incorporating the leaked sidemode states into the prepared states. In this way, the leaked state information can be faithfully incorporated to the security frame and attributed to the insecure quantum channel.

\medskip
\noindent\textbf{Data availability} \\
The data that support the findings of this study are available from the corresponding author upon reasonable request.\\

\end{document}